# Phosphorene Oxide: Stability and electronic properties of a novel 2D material

Gaoxue Wang[1], Ravindra Pandey[1*], and Shashi P. Karna[2]

[1]Department of Physics, Michigan Technological University, Houghton, Michigan 49931, USA
[2]US Army Research Laboratory, Weapons and Materials Research Directorate, ATTN: RDRL-WM, Aberdeen Proving Ground, MD 21005-5069, U.S.A.

(October 22, 2014)

*Email: gaoxuew@mtu.edu,
pandey@mtu.edu
shashi.p.karna.civ@mail.mil

**Abstract**

Phosphorene, the monolayer form of the (black) phosphorus, was recently exfoliated from its bulk counterpart. Phosphorene oxide, by analogy to graphene oxide, is expected to have novel chemical and electronic properties, and may provide an alternative route to synthesis of phosphorene. In this letter, we investigate physical and chemical properties of the phosphorene oxide including its formation by the oxygen adsorption on the bare phosphorene. Analysis of the phonon dispersion curves finds stoichiometric and non-stoichiometric oxide configurations to be stable at ambient conditions, thus suggesting that the oxygen absorption may not degrade the phosphorene. The nature of the band gap of the oxides depends on the degree of the functionalization of phosphorene; indirect gap is predicted for the non-stoichiometric configurations whereas a direct gap is predicted for the stoichiometric oxide. Application of the mechanical strain and external electric field leads to tunability of the band gap of the phosphorene oxide. In contrast to the case of the bare phosphorene, dependence of the diode-like asymmetric current-voltage response on the degree of stoichiometry is predicted for the phosphorene oxide.

# 1.0 Introduction

Two dimensional (2D) materials have been extensively studied due to their novel properties and technological important applications. Especially, the discovery of graphene has stimulated an avalanche of investigations to exploit its novel properties for applications at nanoscale [1]. In the post-silicon era, graphene has been widely regarded as the most promising building blocks for the electronic devices. However, its metallic nature together with sensitivity to the environment leads to somewhat limited scope of applications. A finite band gap in a material is known to be essential for the fabrication of device elements including transistors. Such a limitation associated with graphene has led to exploration of 2D materials beyond graphene [2].

Phosphorene has recently attracted attention as an alternative to graphene due to its direct band gap of 2 V and a large current on/off ratio with the hole mobility of about 1000 $cm^2V^{-1}s^{-1}$ [3-5]. Fabrication of the phosphorene-based optical devices with broadband photo responses has recently been reported [6, 7]. Note that phosphorene is successfully exfoliated from the most stable allotrope of phosphorus, black phosphorus [3, 8, 9].

The chemical modification of 2D materials has now routinely been performed to tailor their physical, chemical and electronic properties. In the case of graphene, surface modifications by H, O, and F atoms often lead to substantial changes in its electronic structure. For example, H- and F-functionalized graphene are wide band gap materials, whereas graphene has zero-gap at the Dirac point [10-12]. Also, graphene oxides are the structures with the presence of the oxygen functional groups on graphene which show remarkable mechanical strength and tunable optoelectronic properties and have broadly used for large scale fabrication of graphene [13-16]. Furthermore, a recent paper argues that the oxidation of phosphorene will result in its degradation by the formation of molecular compounds such as phosphorus trioxide ($P_4O_6$), and phosphorus pentoxide ($P_4O_{10}$) or phosphates [17, 18]. However, such conjecture has not yet been verified by experimental or theoretical studies.

In this letter, we present the results of our theoretical study based on density functional theory on the oxide configurations of phosphorene. We first look into the interaction of oxygen in both atomic and molecular form with the bare phosphorene predicting the preferred binding site and energy barrier to dissociate the oxygen molecule adsorbed on the monolayer. The effect of degree of the oxygen functionalization on the stability and electronic structure of phosphorene will be examined. We also will calculate the transverse electron transport properties of the oxide configurations in a model setup mimicking the Scanning Tunneling Microscopy (STM) experiment.

**2.0 Computational details**

Electronic structure calculations were performed by the density functional theory (DFT) method using the norm-conserving Troullier-Martins pseudopotential as implemented in SIESTA [19]. The Perdew-Burke-Ernzerhof (PBE) [20] exchange correlation functional was employed. A double-zeta basis including polarization orbitals was used. The energy convergence was set to $10^{-5}$ eV. The mesh cutoff energy was chosen to be 500 Ry. The geometry optimization was considered to be converged when the residual force on each atom was smaller than 0.01 eV/Å. In our periodic supercell approach, the vacuum distance normal to the plane was larger than 20 Å to eliminate interaction between the replicas. A dipole correction was employed to eliminate the artificial electrostatic field between the periodic supercells.

For calculations describing interaction of an O atom and an $O_2$ molecule with phosphorene, a (3×4) supercell with a total of 48 P atoms was used and the reciprocal space was sampled by a grid of (5×5×1) $k$ points in the Brillouin zone. On the other hand, PO, $P_2O_1$ and $P_4O_1$ configurations were calculated with a (1×1) supercell consisting of 4 P atoms, and the $P_8O_1$ configuration was calculated with a (1×2) supercell consisting of 8 P atoms. The $k$-point mesh of (11×11×1) was used for these oxide configurations. The phonon dispersion calculation was based on Vibra of SIESTA utility [21].

Phosphorene has a puckered surface due to the sp$^3$ character of the chemical bonds at the surface. We find the bond lengths and the bond angles to be (2.29, 2.26 Å), (103.7°, 95.6°) which are in agreement with the previously reported values obtained at the PBE-DFT level of theory [22, 23]. Likewise, calculations using the same modeling elements reproduced the structural and electronic properties of graphene-based systems [24-27], thereby showing accuracy and reliability of our computational model in describing 2D materials.

**3.0 Results and discussion**

*3.1 Monoatomic Oxygen (O)*

First, we investigate the interaction of a single oxygen atom with phosphorene. Figure 1 shows the lattice sites considered for the oxygen approaching phosphorene e.g. (i) the ring site - above the center of hexagonal ring, (ii) the top site - above the top of P atom, and (iii) the bridge site - above the bridge of P-P bonds. Interestingly, O atoms approaching either top or ring sites prefer the same equivalent positions in their equilibrium configurations having tetrahedral coordination for P atoms.

In the equilibrium configuration, $R_{(P-O)}$ is ~1.54 Å, and the binding energy/oxygen atom defined as $(E_{(Phosphorene)}+E_{(O2\ molecule)})/2 - (E_{(Phosphorene+O\ atom)})$ is calculated to be 1.4 eV/oxygen atom. The bridge site configuration is ~2.3 eV higher in energy than the equilibrium configuration which is in contrast to the case of graphene oxide where the bridge site is found to be the preferred binding site for oxygen [16]. The adsorbed oxygen atom does not induce either magnetism or mid-gap states as shown by the spin polarized density of states (DOS), see supplementary information, Fig S1.

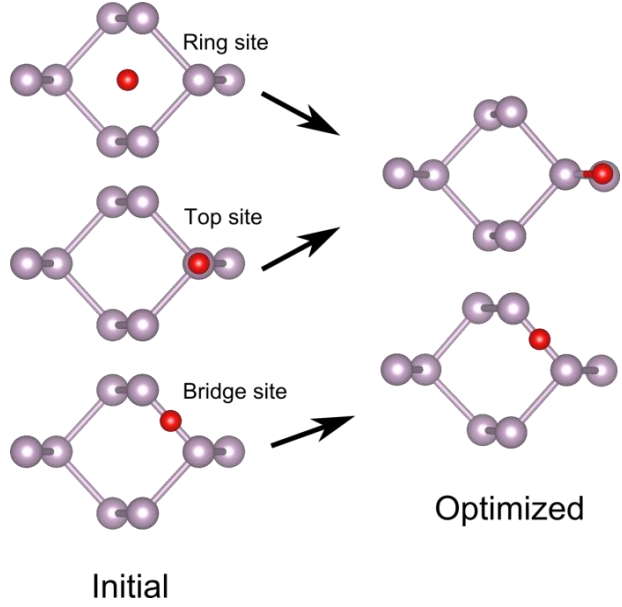

*Figure 1: Single oxygen atom absorption on phosphorene. The left panel shows the initial configurations and the right panel shows the optimized configurations. The oxygen atoms are in red, and phosphorus atoms in purple.*

### *3.2 Molecular Oxygen ($O_2$)*

Next, we investigate the interaction of an oxygen molecule with phosphorene considering both adsorption and dissociation processes on the surface. Figure 2(a) shows the calculated ground state configuration of the adsorbed oxygen molecule. Here, $O_2$ prefers a tilted orientation with $R_{(P-O1)}$=1.69 Å and $R_{(P-O2)}$=1.75 Å. Note that the adsorbed $O_2$ gets stretched out on the surface with $R_{(O-O)}$ of 1.60, which is substantially larger than that of $O_2$ (~1.24 Å). The binding energy defined as ($E_{(Phosphorene)}+E_{(O2\ molecule)} - E_{(Phosphorene+O2\ molecule)}$) is found to be about 0.78 eV/oxygen molecule. The binding energy of an oxygen molecule is smaller than the binding energy of two separated oxygen atoms due to the energy needed to stretch the $R_{(O-O)}$ to 1.60 Å. Note that the spin polarization was considered for the binding energy calculations.

The dissociation process of the adsorbed $O_2$ is simulated by fixing an O atom (i.e. O1), and moving the other atom (i.e. O2) laterally in the unit cell (shadowed region) as shown in the inset of Figure 2(a). A minimum occurs in the corresponding energy surface as shown in Figure 2(b) when O2 moves toward the P2 site. The calculated energy barrier is 0.33 eV along the path illustrated by the arrows in the inset, Figure 2(c). The energy barrier increases to 1.31 eV when we move O2 to a new atomic site, P3. In an alternative scenario of moving the O1 atom and fixing the O2 atom, a much higher energy barrier is expected since the O1 atom is strongly bonded to P atom with the bond distance of 1.03 Å (see supplementary information, Fig S2). Note that the dissociation energy of an $O_2$ molecule on the bare graphene is about 2.39 eV [28], which is much larger than the dissociation energy of the $O_2$ molecule on phosphorene (~0.3 eV). Dissociation of the adsorbed $O_2$ on phosphorene can therefore be one of the possible chemical routes to form the phosphorene oxide [18].

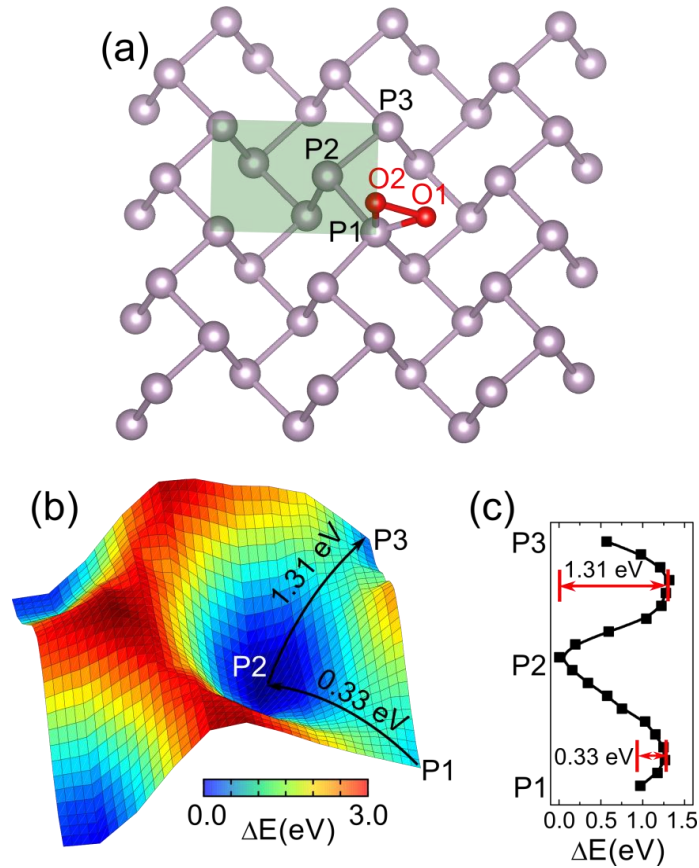

*Figure 2. $O_2$ on phosphorene: (a) the ground state configuration, (b) the energy surface showing displacement of an O atom from P1 to P2 to P3 atomic sites, and (c) the calculated energy barrier along the paths as shown by the arrows in (b). The oxygen atoms are in red, and phosphorus atoms in purple.*

## 3.3. Stoichiometric phosphorene oxide (PO)

Phosphorene has a puckered surface (Figure 3(a)), and addition of an O atom at each atomic site leads to a configuration of PO with a slight increase in the P-P bond length (2.32, 2.37 Å) as compared those for the bare phosphorene. The length of P-O bond is 1.51 Å (Figure 3(b)) which is similar to the C-O bond length of 1.47 Å in graphene oxide [29], and is slightly larger than the distance between B-O of 1.40 Å for $O_2$ adsorbed on the $B_{13}$ cluster [30].

As seen from the side view of Figure 3(b), PO is deformed compared to the bare phosphorene with changes in bond angles between P atoms. However, the structure retains its original configuration without cleavage of P-P bonds. This is different from the cases of H, F, and -OH absorption which act as chemical scissors and break down phosphorene into nanoribbons [31].

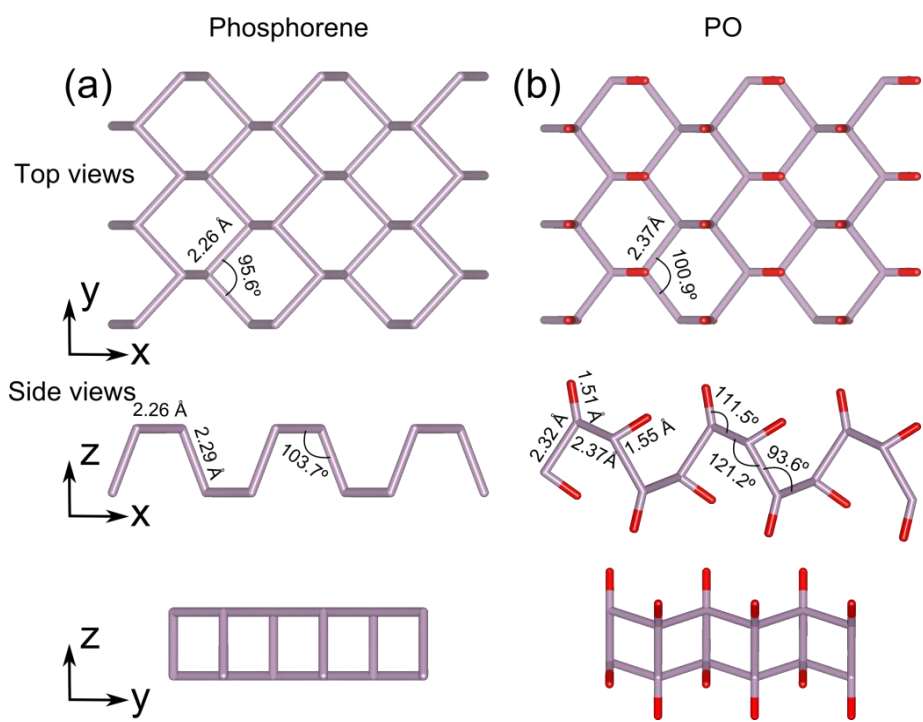

Figure 3. Top and side views of (a) phosphorene, and (b) phosphorene oxide.

In the stoichiometric PO configuration, the ∠P-P-P bond angles are 121.2°, 93.6° and 100.9°. The change in the ∠P-P-P bond angles, relative to the bare phosphorene, is closely related to the charge redistribution as shown in the deformed charge density plot, see supplementary information, Fig S3. Analysis of the Mulliken charges finds that ~0.2 e is transferred from a P atom to an O atom. This is different from the case of graphene oxide where O tends to bind on the bridge of C-C bond forming epoxy groups, and the accumulated electron density around O comes from two neighboring C atoms of graphene [29, 32, 33]

The stability of the phosphorene oxide is confirmed by the calculated phonon dispersion curves showing no negative frequencies (Figure 4). The phonon dispersion of PO is greatly different from that of phosphorene; it can be grouped into three regions (Figure 4(b)) with the highest vibrational frequency of about ~1160 cm$^{-1}$. On the other hand, the phonon dispersion curves of the bare phosphorene have separated acoustic and optical modes with the maximum vibrational frequency of ~460 cm$^{-1}$. In the lower acoustic region for PO, the vibrational modes are associated with the constituent P and O atoms. The modes associated with the P atoms dominate in the middle region of the spectrum. The high frequency modes correspond to the P-O stretching modes indicating a relative high strength of the P-O bond in the 2D lattice.

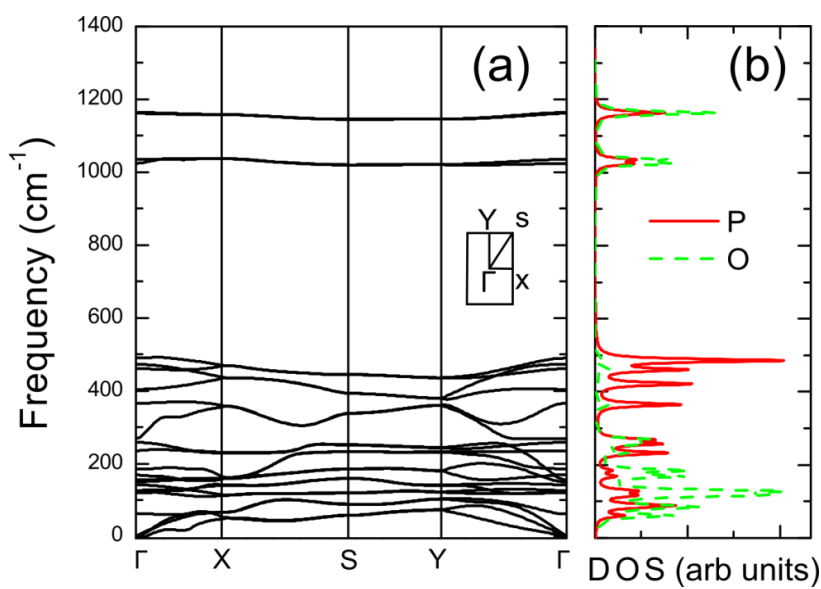

*Figure 4. Phosphorene oxide (a) the phonon dispersion curves, and (b) density of states.*

In our study, we consider the oxygen adsorption on phosphorene to be more-like an ordered absorption of adsorbates on graphene, such as the case of graphane, fluorographene, and chlorographene [10, 12, 34]. In graphene oxide, the oxygen functional groups form a inhomogeneous lattice as revealed by the transmission electron microscopy (TEM) measurements [33]. This may be due to interaction of oxygen atoms at the top and bottom sides of graphene leading to clustering of oxygen atoms. A recent theoretical study predicts the formation of ordered, homogenous of single surface graphene oxides by oxidizing only the top layer of graphene [16]. A formation of the homogenous graphene oxides is also observed after the oxidation of epitaxial graphene grown on a SiC substrate [35].

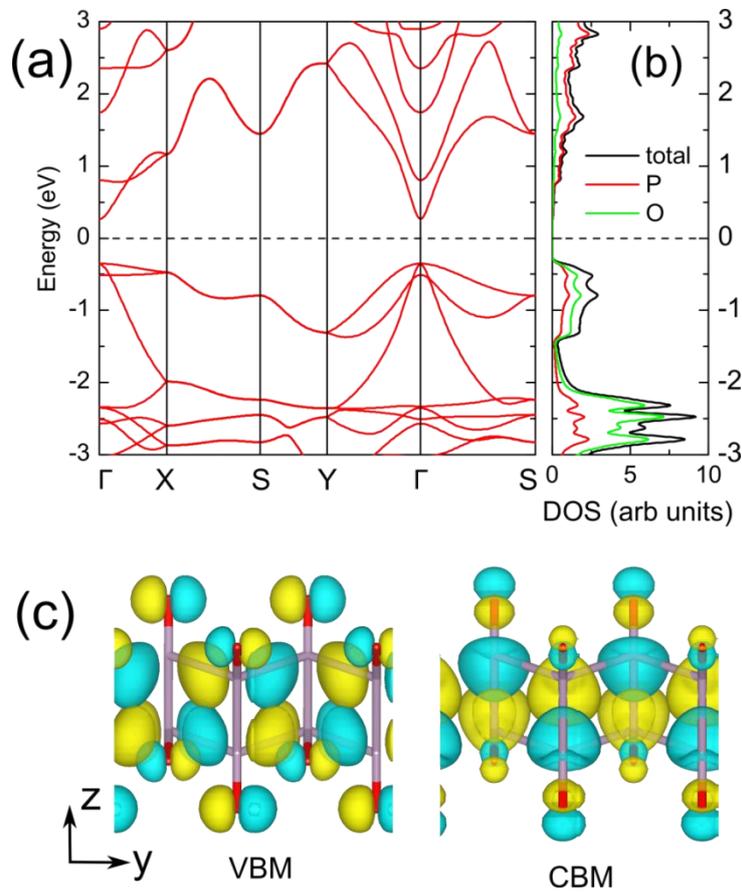

*Figure 5. Electronic properties of phosphorene oxide: (a) band structure, (b) density of states, and (c) Kohn-Sham wave functions at $\Gamma$ associated with states corresponding to top of the valence band (VBM) and bottom of the conduction band (CBM). $\Gamma$, S, Y and X in the k-space are defined as (0,0,0), (1/2,1/2,0), (0,1/2,0), and (1/2,0,0), respectively.*

The calculated band structure of PO is shown in Figure 5. The valence band maximum (VBM) has $p_y$ character associated with both P and O atoms (Figure 5(c)), and the conduction band minimum (CBM) is formed by P-$s$ orbitals and O-$p_z$ orbitals (Figure 5(c)).The calculated band gap is direct at $\Gamma$ with a value of ~0.6 eV. It is smaller in magnitude than that calculated for the bare phosphorene (~1 eV at the PBE-DFT level of theory). Anisotropy in the band structure of the 2D lattice is predicted; the hole effective masses are 4.56 and 1.74 $m_e$ along $\Gamma$-X and $\Gamma$-Y directions, respectively. On the other hand, the electron effective mass does not show anisotropy and has a magnitude of 0.18 $m_e$ (see supplementary information, Fig S4).

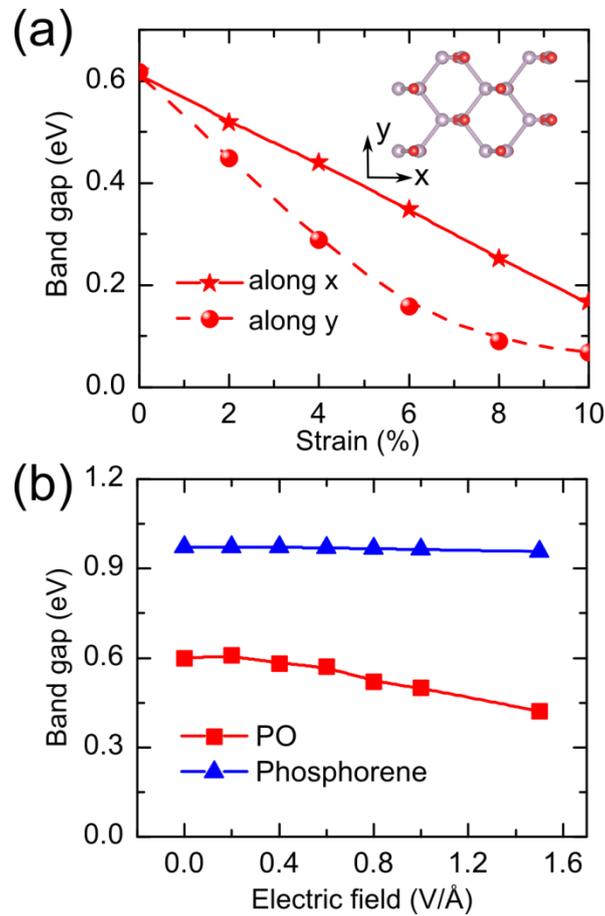

*Figure 6. Phosphorene oxide: Band gap vs. in-plane tensile strain, and (b) band gap vs. electric field applied perpendicular to the 2D lattice.*

Application of the tensile strain along *x* yields a linear variation of the band gap in a range of values (0.1-0.6 eV) for the strain values of 0% to 8% (Figure 6(a)). The predicted variation in the band gap is mainly due to variation of the conduction band minimum (CBM) at $\Gamma$ (see supplementary information, Figs. S5 and S6). The top of the valence band (VBM) which is mainly formed by the oxygen atoms does not appear to be sensitive to the external strain (see supplementary information, Fig S6). Likewise, the external electric field modifies the band gap reducing it to be 0.4 eV at 1.5 V/Å, while it does not change the band gap for the bare phosphorene (Figure 6(b)). This is due to the reduced symmetry in the oxide configuration compared to its bare configuration. Thus, the oxygen functionalization of phosphorene yields tunability of the band gap with both tensile strain and electric field.

The in-plane stiffness of the 2D lattice is calculated by fitting the strain energy within the strain range of -2% to 2% (see supplementary information, Fig S7) [24]. The calculated stiffness constant for phosphorene is 21 and 66 N/m along *x* and *y* directions, respectively. For the phosphorene oxide, the stiffness constant is decreased to 16 and 33 N/m along *x* and *y* directions due to the increased P-P bond length in the oxide lattice. These values are much smaller those associated with graphene (340 N/m [36]) which suggest the softness of phosphorene-based 2D materials [37].

*3.4 Non-stoichiometric phosphorene oxide*

We now investigate stabilities and electronic properties of non-stoichiometric phosphorene oxides representing the cases of the partial functionalization of the phosphorene. Figure 7 shows the considered non-stoichiometric oxide configurations including $P_8O_1$ (i.e. $PO_{0.125}$), $P_4O_1$ (i.e. $PO_{0.25}$), and $P_2O_1$ (i.e. $PO_{0.5}$), with a single side absorption of the oxygen atoms.

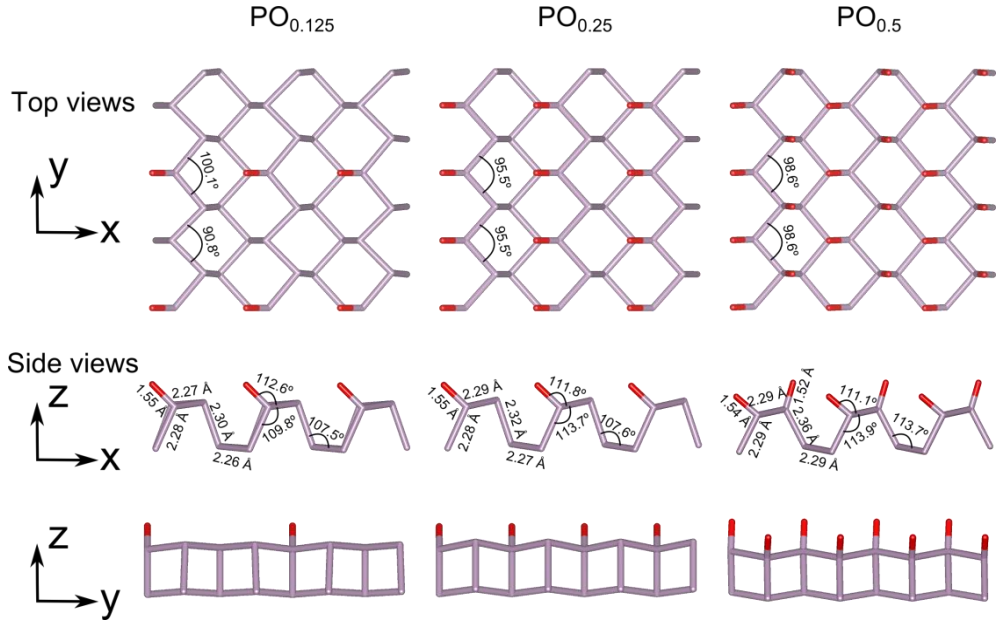

*Figure 7. Structures of non-stoichiometric oxide configurations: $PO_{0.125}$, $PO_{0.25}$, and $PO_{0.5}$.*

Figure 8 shows the variation of the band gap with the degree of the functionalization of phosphorene. Bare phosphorene is a direct gap 2D material. This is not the case with phosphorene oxides where the nature of the band gap depends on the degree of functionalization; an indirect band gap is predicted for $PO_{0.125}$, $PO_{0.25}$, and $PO_{0.5}$. Finally, a crossover from indirect to direct band gap is seen for the stoichiometric PO configuration. The direct band gap is defined as the energy gap at Γ. Since the location of VBM and CBM depends on the degree of functionalization of phosphorene (see supplementary information, Fig. S8), the indirect band gap is defined as (Γ→Γ-Y), (Γ→Γ-X), (S→Y), (Γ →Γ-X), and (Γ→Γ-X) for P, $PO_{0.125}$, $PO_{0.25}$, $PO_{0.5}$ and PO, respectively.

The phonon dispersion curves of non-stoichiometric configurations, see supplementary information, Fig S8, suggest the stability of the partially oxidized configurations, though small negative values (~8 cm$^{-1}$) near Γ may be an artifact of the numerical approximation employed in the frequency calculations.

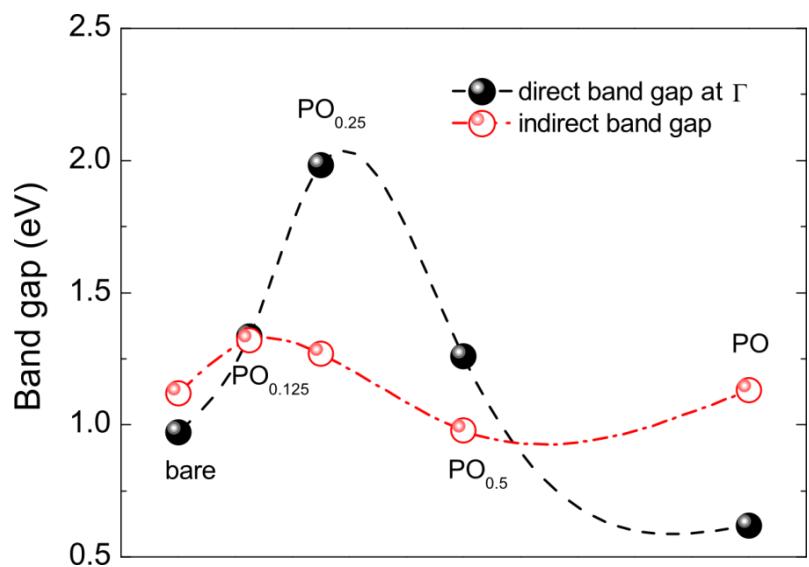

*Figure 8. The variation of band gap as a degree of functionalization of the bare phosphorene. Open and solid circles represent the values of indirect and direct band gaps, respectively. The direct band gap is taken to be at $\Gamma$.*

The work function is a crucial physical quantity to determine the emission properties of materials and have considerable impact on device performance. It is defined as energy difference between the vacuum level and Fermi energy. For the oxide configurations, the work function shows a monotonous increase with the increased degree of oxygen functionalization of the bare phosphorene. This is due to the fact that the charge transfer from P to O as shown by the deformation charge density plot (see supplementary information, Fig S3) will lead to formation of dipoles between the phosphorus layer and oxygen layer, thus preventing electrons moving toward the vacuum. The calculated values of the work function for bare phosphorene, $PO_{0.125}$, $PO_{0.25}$, $PO_{0.5}$, and PO are 4.5, 4.9, 5.2, 5.8, and 7.2 eV, respectively. Therefore, the work function can be tailored effectively with the degree of oxygen functionalization of phosphorene. Similar tunable work function has already been reported for graphene; the work function increases from 4.2 eV to 5.5 eV with 20% concentration of oxygen functionalization [38, 39].

## 3.5 Tunneling characteristics

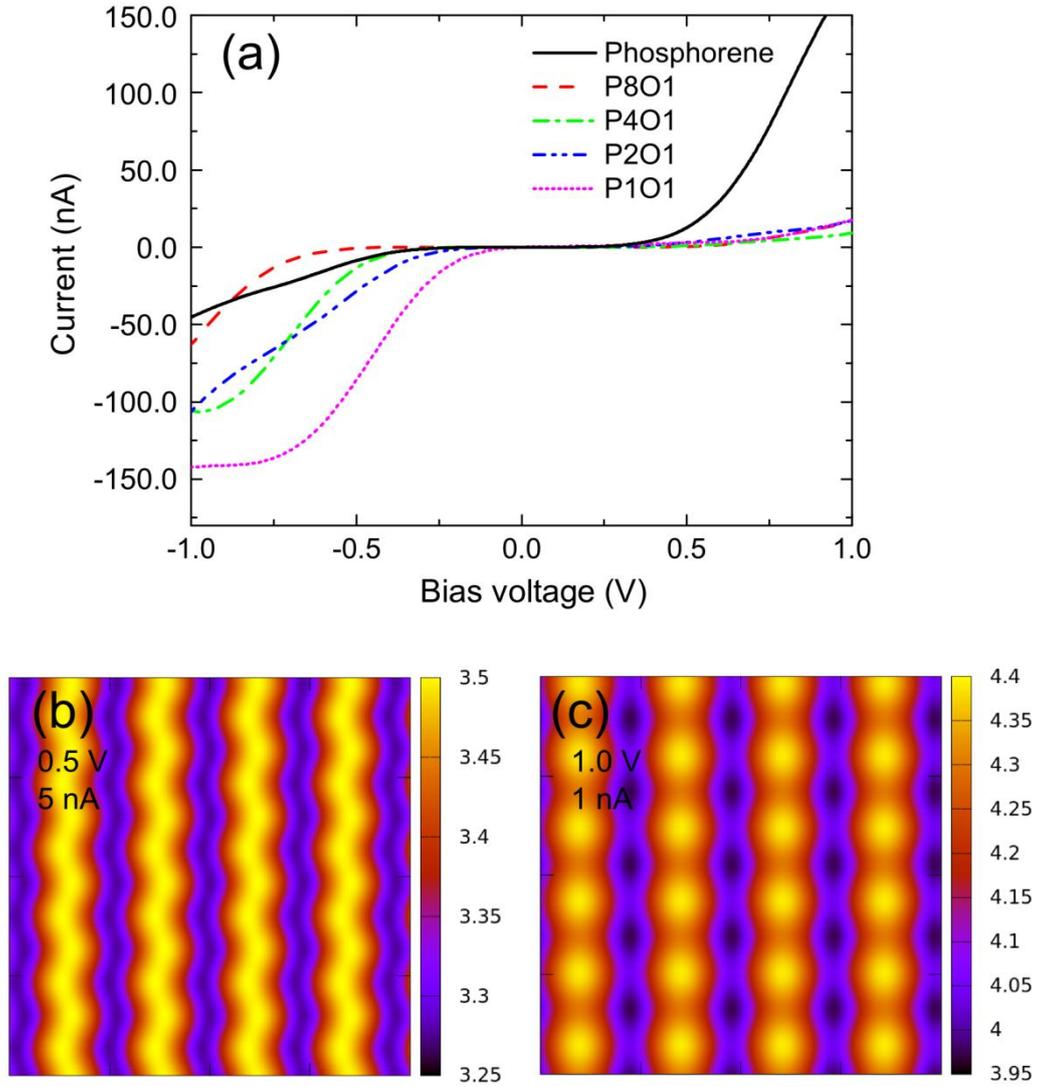

*Figure 9. (a) Tunneling characteristics of the phosphorene oxide configurations. (b) The simulated STM images of phosphorene and the phosphorene oxide. The current is calculated using a Au13 tip located at 3 Å above the surface. The side scale bar shows the distance from the tip to the surface in unit of angstrom.*

Finally, the tunneling characteristics of the phosphorene oxides are investigated. The tunneling current from the sample to the tip at location $\vec{r}_t$ based on Tersoff and Hamann approximation [40] is

$$I(\vec{r}_t; V) \approx \frac{2\pi e}{\hbar} \int_{-\infty}^{+\infty} \rho_t\left(E - \frac{eV}{2}\right) \rho_s\left(\vec{r}_t; E + \frac{eV}{2}\right) F(E) dE \qquad (1)$$

where $\rho_t$ is the electron density of the tip, $\rho_s$ is the electron density of the sample at the location of the tip. $F(E)$ is the term to include the effect of thermally excited electrons as proposed by He *et al*. [41-44] In order to mimic the scanning tunneling micriscope (STM) measurements, we use the constant current mode with the gold tip represented by a $Au_{13}$ cluster. The size of the STM images are (20 Å×20 Å), and a positive bias between the sample and the tip was applied. It is to be noted here that a positive bias between the sample and the tip will lead to tunneling of electrons from tip-VB to sample-CB. With a negative bias, the electrons will tunnel from sample-VB to tip-CB (see supplementary information, Fig S9).

The electron transport studies offer some intriguing insight into electron tunneling in the direction perpendicular to the stable phosphorene oxide plane. Asymmetric current-voltage (IV) semiconducting characteristics (Figure 9(a)) are seen for both stoichiometric and non-stoichiometric oxide configurations. Bare phosphorene shows strong diode like behavior with large tunneling current in the positive bias regime. Upon oxygen adsorption, this behavior completely changes, with high current in the negative bias regime. The high tunneling current of PO in the negative bias region is due to the contribution of oxygen atoms in the top of valance band (Figure 5(b)).

The dependence of the threshold voltage onset of the tunneling current on the degree of functionalization suggests that tunable electronic properties can be achieved by the oxygen functionalization of phosphorene. In the negative bias region, the threshold voltage decreases from -0.55 V ($P_8O_1$) to -0.15 V (PO) which is related to variation in the band gap of oxides. Furthermore, the STM images as seen in Figures 9(b) and (c) can help in identifying formation of the phosphorene oxide from the bare phosphorene.

**Summary**

The interaction of phosphorene with oxygen and the formation of 2D phosphorene oxide were investigated with the use of the density functional theory. A number of key findings have emerged from this study based on density functional theory. First, our calculations predict that the 2D phosphorene oxide to be stable in both stoichiometric and non-stoichiometric configurations. Second, a fully functionalized phosphorene is a direct band gap material with tunable band gaps by external strain and electric field. Partially functionalized phosphorene has an indirect band gap. Third, the dissociation energy of an oxygen molecule is calculated to be ~0.33 eV, suggesting possible, low-energy oxidation of phosphorene which is likely to lead to the 2D phosphorene-based structures. Finally, electron transport studies offer some intriguing insight into electron tunneling in the direction perpendicular to the phosphorene oxide plane with dependence of the current on the degree of functionalization of phosphorene. We believe that the results of this study would inspire experimental efforts into the synthesis and electronic device physics studies of phosphorene oxide.

**Acknowledgements**

The authors thank D. R. Banyai for providing his STM simulation code, and appreciate Dr. S. Gowtham for his support on the installation of required software. RAMA and Superior, high performance computing clusters at Michigan Technological University, were used in obtaining results presented in this paper. Financial support from ARL W911NF-14-2-0088 is obtained.

**Supplementary Material:**

**"Phosphorene Oxide: Stability and electronic properties of a novel 2D material"**
Gaoxue Wang, Ravindra Pandey, and Shashi P. Karna

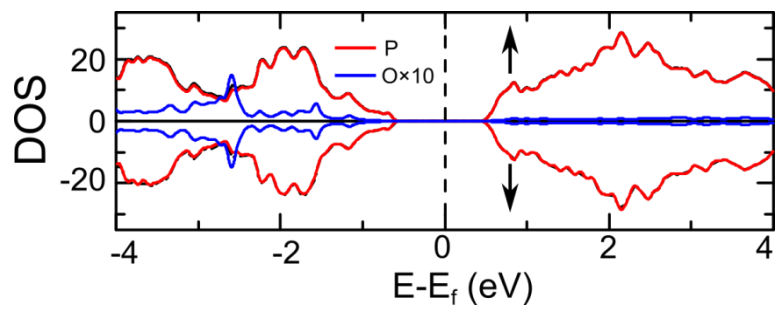

*Figure S1. Density of states with a single oxygen adsorbed on phosphorene.*

In the equilibrium configuration (Figure S2), $O_2$ binds to a phosphorus atom (P1) with a tilt configuration where $R_{(P1-O1)} = 1.69$ Å and $R_{(P1-O2)} = 1.75$ Å. As seen from the side view, O1 is closer to the surface with the distance of $d=1.03$ Å whereas O2 is far away with from the surface with the distance of about 1.75 Å. Therefore, O2 was moved laterally to determine the dissociation energy. A lateral displacement of O1 atom while fixing O2 atom will lead to a higher energy barrier since the distance of O1 from the top P atoms is only 1.03 Å.

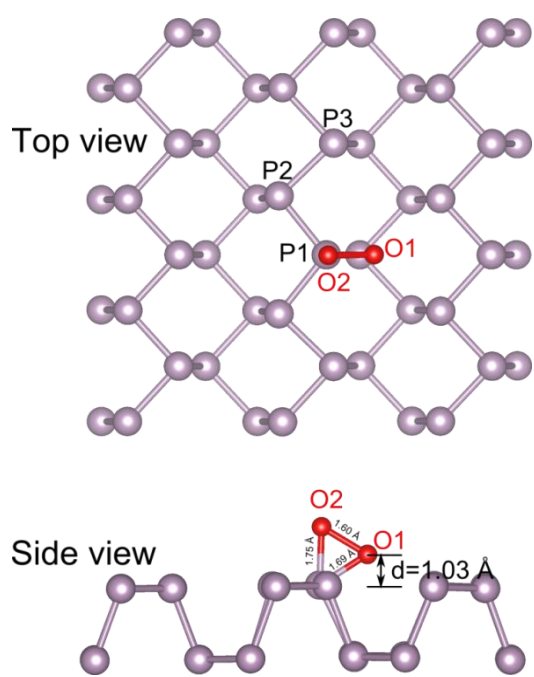

Figure S2. The top and side views of an oxygen molecule adsorbed on the phosphorene surface.

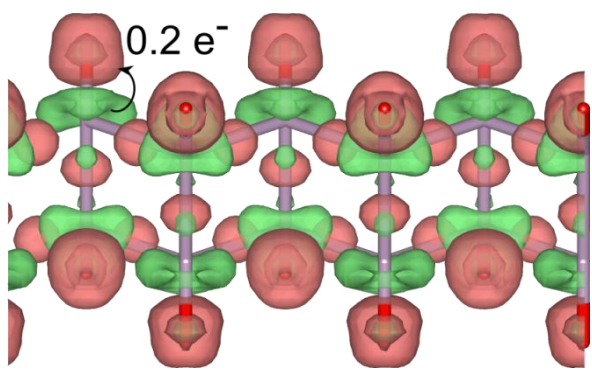

*Figure S3. Deformation charge density of phosphorene oxide. The red region represents the accumulation of electrons and green region represents the depletion of electrons.*

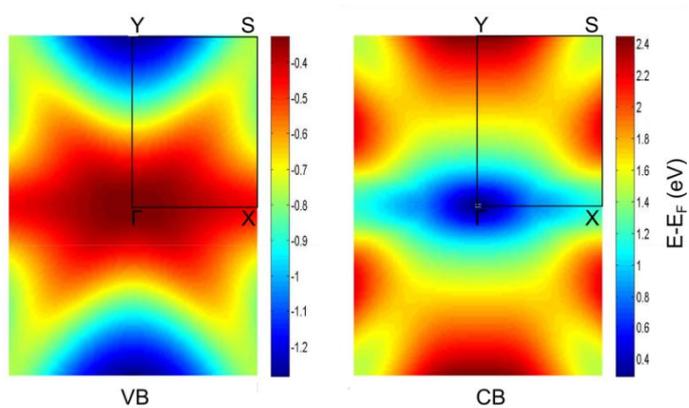

*Figure S4. Two dimensional band structure of phosphorene oxide: (a) valence band, (b) conduction band.*

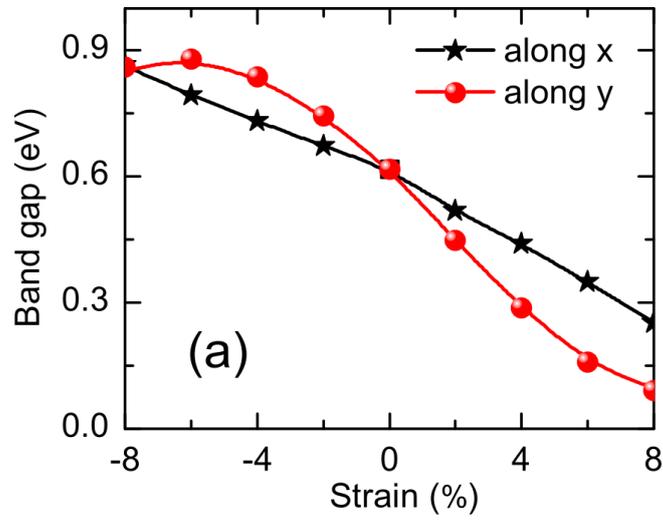

*Figure S5. In-plane compressive and tensile strain effect on the band gap of phosphorene oxide.*

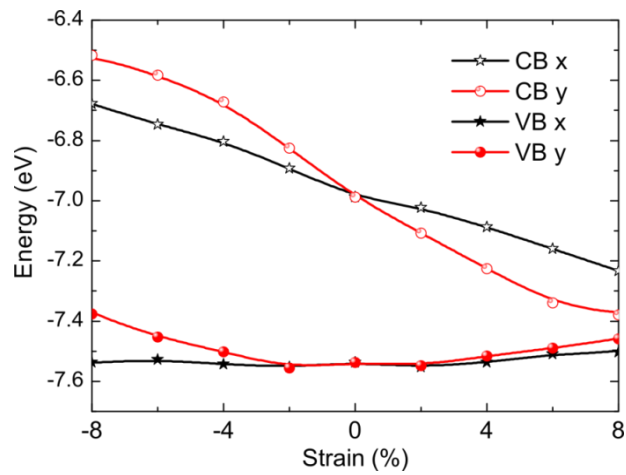

*Figure S6. Phosphorene oxide: variation of valence and conduction bands at Γ point with strain along x (black) and y (red) directions: top of the valence band (solid dots) and bottom of the conductive band (open dots)*

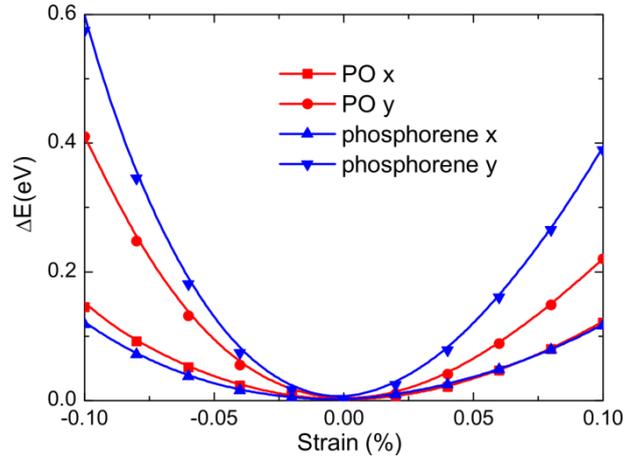

*Figure S7. Variation of energy with strain for phosphorene and phosphorene oxide.*

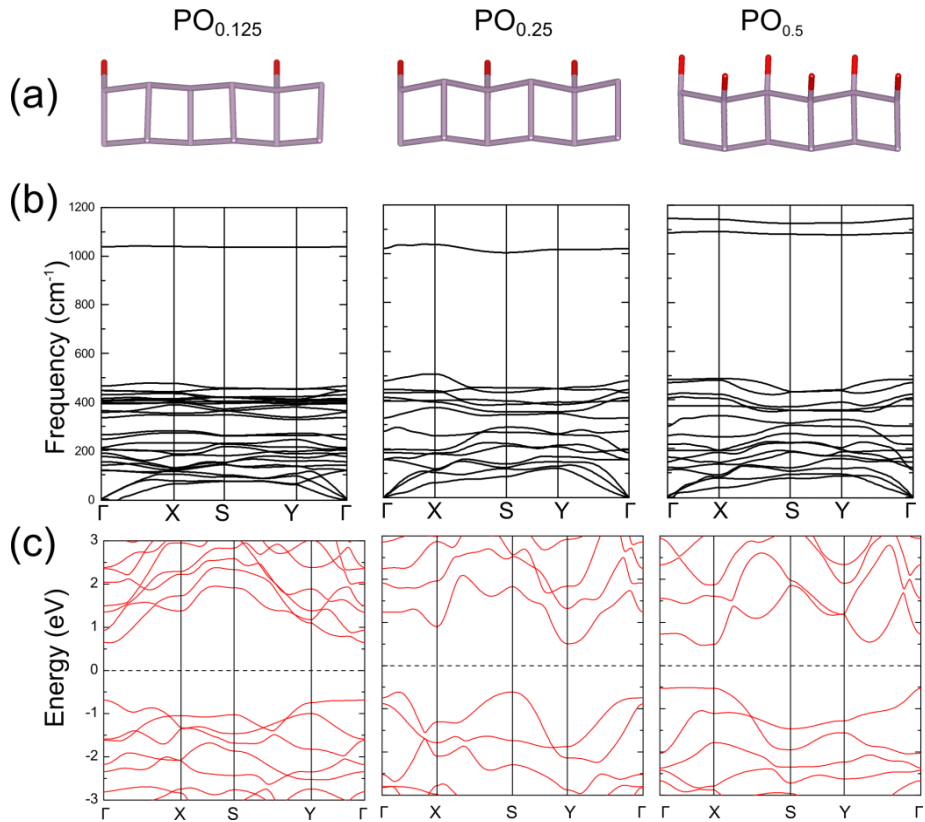

*Figure S8. (a) Structure, (b) phonon dispersion curves, and (c) band structures for $PO_{0.125}$, $PO_{0.25}$, and $PO_{0.5}$. The direct band gap is defined as the energy gap at $\Gamma$. The indirect band gap is defined as $(\Gamma \rightarrow \Gamma\text{-}X)$, $(S \rightarrow Y)$, and $(\Gamma \rightarrow \Gamma\text{-}X)$, and $(\Gamma \rightarrow \Gamma\text{-}X)$ for $PO_{0.125}$, $PO_{0.25}$, $PO_{0.5}$, respectively. $\Gamma$, S, Y and X are defined as (0,0,0), (1/2,1/2,0), (0,1/2,0), and (1/2,0,0), respectively.*

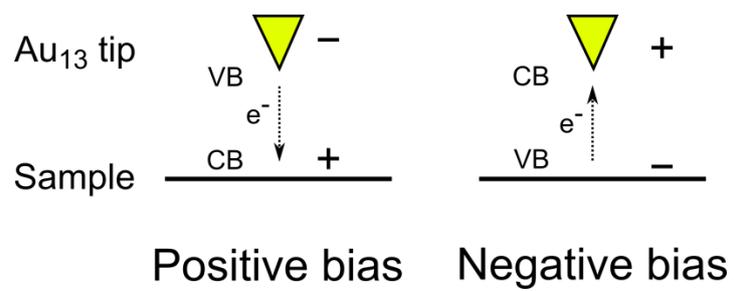

*Figure S9. Schematic illustration of the STM set up.*